\documentclass[prb,preprintnumbers,superscriptaddress,amsmath,amssymb,twocolumn]{revtex4}

\usepackage{graphicx}
\usepackage{dcolumn}
\usepackage{bm}
\usepackage{latexsym,epsfig}
\usepackage{chemarrow}
\usepackage{xcolor}
\begin{document}

\title{Connecting the one-band and three-band Hubbard models of cuprates via spectroscopy 
and scattering experiments}

\author{K. Sheshadri} 

\affiliation{226, Bagalur, Bangalore North, Karnataka State, India 562149 }

\author{D. Malterre} 

\affiliation{Institut Jean Lamour, Universit\'{e} de Lorraine, UMR 7198 CNRS, BP70239, 54506 Vandoeuvre l\'{e}s Nancy, France}

\author{A. Fujimori}
\affiliation{Department of Physics, The University of Tokyo, 7-3-1 Hongo, Bunkyo-ku, Tokyo 113-0033, Japan}
\affiliation{Center for Quantum Science and Technology and Department of Physics, National Tsing Hua University, Hsinchu 30013, Taiwan}
\affiliation{Condensed Matter Physics Group, National Synchrotron Radiation Research Center, Hsinchu 30076, Taiwan }

\author{A. Chainani} 

\affiliation{Condensed Matter Physics Group, National Synchrotron Radiation Research Center, Hsinchu 30076, Taiwan }

\date{\today}

\begin{abstract}

The one-band and three-band Hubbard models which describe the electronic structure of cuprates  
indicate very different values of effective electronic parameters, such 
as the on-site Coulomb energy and the hybridization strength. 
In contrast, a comparison of electronic parameters of
several cuprates with corresponding values from spectroscopy and scattering experiments
indicates similar values in the three-band model and
cluster model calculations used to simulate experimental results. 
The Heisenberg exchange coupling $J$ obtained by a downfolding method in 
terms of the three band parameters is used to carry out  
an optimization analysis consistent with $J$ from neutron scattering experiments for a series of cuprates. In addition, the effective one-band parameters $\tilde{U}$ and $\tilde{t}$ are described using the three band parameters, thus revealing the hidden equivalence of the one-band and three-band models. The ground-state singlet weights obtained from an exact diagonalization
elucidates the role of Zhang-Rice singlets in the equivalence. The results provide a consistent method to connect electronic parameters obtained from spectroscopy and the three-band model with values of $J$ obtained from scattering experiments, band dispersion measurements and the effective one-band Hubbard model.  

\end{abstract}

\maketitle

\section{Introduction}
\label{introduce}

The mechanism of high-temperature superconductivity exhibited by the copper-oxide families of layered compounds remains one of the most intriguing and challenging topics in condensed matter physics \cite{Keimer}, nearly 36 years after its discovery \cite{Bednorz}. 
The discovery of copper-oxide based superconductivity led to unprecedented theoretical and experimental efforts to understand the phenomenon. While there have been innumerable models put forth to understand the mechanism of high-temperature superconductivity, it still remains an open problem. On the other hand, nearly all the models agree that superconductivity
in the copper-oxide based materials is intimately associated with quasi two-dimensionality (2D) and strong electron-electron correlations \cite{Keimer}. This is based on the fact that the CuO$_2$ planes are the main source of the electronic states which undergo the superconducting transition. At a very broad level,  
the possible mechanisms discussed in the literature span over various models including the effective one-band Hubbard model \cite{Anderson,Zhang}, resonating valence bond theory \cite{Baskaran}, the three-band Hubbard model \cite{Varma01, Emery}, the t-J model \cite{Spalek}, spin-fluctuation driven pairing \cite{Schulz}, marginal Fermi liquid \cite{Varma02}, pair density wave model \cite{Agterberg}, electron-phonon coupling-induced pairing which go beyond the BCS model \cite{Bishop}, and so on. 

The simplest parent compound La$_2$CuO$_4$ is a good antiferromagnetic insulator and upon hole-doping, undergoes a transition to a dome shaped superconducting phase with an optimal $T_c \sim 38$ K \cite{Bednorz}. 
While the long-range order is lost, La$_{2-x}$Sr$_x$CuO$_4$, as well as several other copper oxide superconductors, continue to exhibit antiferromagnetic  correlations in the form of resonant modes and paramagnons in the superconducting phase 
\cite{Rossat,Eschrig,Tacon}. 
In fact, along with superconductivity, all the families of copper oxide superconductors also show spin- and charge-ordering phenomena\cite{JT,Wu,GG,JC,MT,Tabis,HJ,Julien,Lee,Chan} which suggest a complex coexistence of electron-phonon coupling, spin fluctuations and electron-electron correlation effects \cite{Salkola,Lanzara}.

The basic starting point to understand cuprate properties is often considered to be the 2D Hubbard model involving strong electron-electron correlations with strong Cu-O hybridization leading to the Zhang-Rice singlet (ZRS) ground state\cite{Zhang}.
It is well-known and well-accepted that the parent copper-oxide materials are best described  as charge-transfer insulators in the Zaanen-Sawatzky-Allen scheme \cite{ZSA}, with the copper on-site Coulomb energy $U_d \gg\Delta$, the charge-transfer energy, and the lowest energy excitations involve the strongly hybridized Cu-$3d$ and O-$2p$ ZRS states. Further, hole doping in the parent compound results in oxygen hole carriers retaining the ZRS character of the lowest energy excitations \cite{CTChen,Brookes2001,Brookes}.

A very important issue involves how to quantify electron-electron correlations in any transition metal compound, in general, and the cuprates, in particular \cite{Khomskii,Koch}. Depending on the theoretical model, the values of 
electron-electron correlation strength can be very different for the same material.  A comparison of the effective one band Hubbard model\cite{Anderson,Zhang}  consisting of the single antibonding band made up of the Cu $d_{x^2-y^2}$ and O $p_x$, $p_y$ orbitals, and the effective three band Hubbard model\cite{Varma01,Emery} which describes the cuprate electronic structure in terms of the Cu-O bonding, non-bonding and antibonding bands show significantly different values of on-site Coulomb energy in the Cu $d$ states.
In the following, in order to distinguish the one band and three band parameters, we use the notation $U_d$/$U_p$ and $t_{pd}$ for the Cu/O on-site Coulomb energies and hopping in the three band model, while $\tilde{U}$ and $\tilde{t}$ are used for the one-band Hubbard model, respectively.
Thus, for example, while early studies of the three band model estimated $U_d \sim$ 7-10 eV and $U_p \sim$ 3-6 eV\cite{MM, HSC, ESF}, typical values of $\tilde{U}\sim$ 3-4 eV are known for the effective one-band model \cite{Werner,Hirayama}. 

It is noted that while several theoretical studies have included the oxygen on-site Coulomb energy $U_p$, there are also some studies which have neglected $U_p$. For example, early theory\cite{Emery}, and cluster model calculations of core-level photoemission and optical absorption \cite{Fujimori, ZXS, Mila, Eskes} could explain experimental results fairly well but in the absence of $U_p$.
In a study using the coherent potential approximation, an effective one-band model was obtained from the three band model including the inter-site Coulomb energy $U_{pd}$ treated in the Hartree-Fock approximation, but with $U_p$ = 0 \cite{Aligia01, Aligia02}. The authors showed that the effective  $\tilde{U}$ increased on increasing $U_{pd}$, and they could obtain a metal-insulator phase diagram as a function of $U_d$ and $\Delta$.
In a three-band Hubbard model using the constrained-path Monte Carlo method, 
the binding energy of a pair of holes and the symmetry of superconducting pairing correlation functions was investigated, but in the absence of $U_p$ \cite{Guerrero}. Cluster perturbation theory applied to calculate spectral functions of cuprates also did not include $U_p$, but could show spin-charge separation in the 1D Hubbard model, as well as momentum-dependent spectral-weight in the 2D Hubbard model \cite{Senechal}.
 Cluster dynamical mean field theory approximation was used to investigate the three-band Hubbard model in the absence of $U_p$ and showed that the cuprates can be described as magnetically correlated Mott insulators  \cite{GoMillis}.
More recently, quantum Monte Carlo calculations demonstrated
dynamical stripe correlations in the three-band Hubbard model without $U_p$, and also explained
experimental observations such as the hourglass magnetic dispersion \cite{EHuang}
The three-band Hubbard model using the auxiliary-field quantum Monte Carlo method, but without $U_p$, was used to show the importance of $\Delta$ and a quantum phase transition from an antiferromagnetic insulator to paramagnetic metal for $\Delta$  $<$ 3 eV \cite{Vitali}. 

However, electron spectroscopy studies in conjunction with cluster model calculations or using the Cini-Sawatzky method\cite{Cini,Sawatzky} have estimated $U_d \sim 6-8$ eV \cite{Fujimori,ZXS,Balzarotti} and $U_p \sim 5-6$ eV \cite{Marel,Tjeng,BarDeroma}. Thus, $U_d$ and $U_p$ are comparable and needed for describing the electronic states derived from Cu-O planes, particularly in the-charge transfer limit, as $U_p$ gets close to or larger than $\Delta$.
Further, an {\it ab initio} method with dynamical screening\cite{Werner} applied to the one-band model for La$_{2}$CuO$_{4}$ estimated a static $\tilde{U}(w = 0)$ $\sim 3.65$ eV,  while for the three-band model, it gave $U_d (w = 0)$ $\sim  7.0$ eV and $U_p (w = 0)$ $\sim  4.64 $ eV. Another very recent multi-scale {\it ab initio} method\cite{Hirayama} applied to the one band model for La$_{2}$CuO$_{4}$ estimated $\tilde{U}$ $\sim 5.0$ eV,  while for the three-band model it estimated $U_d \sim  9.6$ eV and $U_p \sim  6.1$ eV.
It is noted that the models have also estimated the inter-site Coulomb energies, as well as the nearest and next-nearest-neighbour hopping ($t$ and $t'$) which also show differences depending on the method \cite{MM, HSC, ESF,Werner,Hirayama}. 

Very interestingly, Hubbard-type cluster models employing $d_{x^2-y^2}$, $p_x$ and $p_y$ levels have been extensively used  for calculating spectra in high energy spectroscopies like core-level photoemission (PES) and x-ray absorption (XAS), resonant inelastic x-ray scattering (RIXS) structure factors, etc. and the obtained electronic parameters\cite{Marel, Tjeng, BarDeroma, NeudertSCO, Okada97SCO, Boske, VeenendaalLCO, Veenendaal02, NCO, NuckerYBCO, Oles, VeenendaalBi2212,Sr2CuO3,Nd2CuO4,YBCO,Dean,Bi2212, Wang,Peng} are quite close to the theoretical estimates from the effective three-band model \cite{MM, HSC, ESF, Johnston,Hirayama} (see Tables I and II). It is noted that the effective three-band model parameters were also used to calculate the dynamical spin structure factor of Bi2201 measured by RIXS\cite{Peng}. On the other hand, analysis of neutron scattering measurements of magnon dispersions\cite{Coldea} and angle-resolved photoemission spectroscopy (ARPES) band dispersions\cite{Leung,Kim} of parent cuprates have employed the extended one-band Hubbard model or the extended $t-J$ model to study the nearest neighbor exchange interaction and correlation effects, and they obtained electronic parameter values close to the values obtained from the effective one-band theoretical models.

For example, it was shown that neutron scattering of La$_2$CuO$_4$ provided a dominant nearest neighbor (NN) hopping  $t$ = 0.33 eV and an effective $U/t$  = 8.8 with $U$ = 2.9 eV, but also showed that in addition to the NN exchange $J$ = 138 meV, it was important to include ring exchange $J_c$ = 38 meV, while $J' = J^{''} = 2$ meV were small\cite{Coldea}. Similarly, for Sr$_2$CuO$_2$Cl$_2$, the
$t-t'-t^{''}-J$ model showed $t$ = 0.35 eV, while $t' = 0.12$ eV, and $t^{''} = 0.08$ eV, and with a
$J = 0.14$ eV \cite{Kim}, it implied an effective $\tilde{U}/t$ = 10 with $\tilde{U}$ = 3.5 eV. Thus, in these cases, the results suggest that the NN hopping
$t$ and $U$ can be considered to be the $\tilde{t}$ and $\tilde{U}$ of the one-band model. For CuO, a recent study 
showed that a linear spin-wave model for a Heisenberg antiferromagnet provided a good description of the neutron scattering results\cite{CuO}. The relevant exchange constants could be accurately determined  and showed that the dominant
exchange interaction $J$ = 91 meV, which coupled antiferromagnetically along the [10$\bar{1}$] chain direction, while the nearest neighbor interchain interactions were very weak ($J_{ac}$ = 3.9 meV  and $J_b$ = 0.39 meV) and coupled ferromagnetically\cite{CuO}.

\onecolumngrid\

\begin{table}[t!]
	\begin{center}
	\caption{Electronic parameters (${U}_d$, $t_{pd}$, ${\Delta}$, ${U}_p$) for cuprates from the three-band Hubbard model (Theory) and from cluster model calculations (Spectroscopy and RIXS). The table also shows two optimized parameter sets ($\Delta_1$ and $U_{p1}$) and ($t_{pd2}$ and $\Delta_2$) with their cost functions f$_1$ and  f$_2$, respectively. $J$ is the nearest-neighbor Heisenberg exchange deduced from scattering experiments. See text for details.}

\begin{tabular}{|c|cccc|ccc|c|ccc|}
\hline
&      &	&	& & & Optimization-1 &	& 	&  &Optimization-2&\\
Compound&      ${U}_d$	&	$t_{pd}$	&	${\Delta}$	& ${U}_p$	&$\Delta_1$ & $U_{p1}$& f$_1$ & $J$(ref.no)		&	$t_{pd2}$	&	$\Delta_2$	& f$_2$ \\

(ref. no) 	&$\pm$0.5 eV	&$\pm$0.2 eV	 &$\pm$1.0 eV&$\pm$0.5 eV& eV & eV & & $\pm$5 meV& eV &	eV &	\\
\hline																									
Theory\\																							
\hline
La$_2$CuO$_4$(34)		&	9.4	&	1.5		&	3.5	&	4.7	&	5.7	&	4.9	&	5.09	&	140 (59)	&	1.1	&	3.7	&	0.96	\\
La$_2$CuO$_4$(35)		&	10.5	&	1.3		&	3.6	&	4.0	&	4.5	&	4.1	&	1.01	&	140 (59)	&	1.2	&	3.7	&	0.4	\\
La$_2$CuO$_4$(36)		&	8.8	&	1.3		&	3.5	&	6.0	&	4.5	&	6.1	&	0.98	&	140 (59)	&	1.2	&	3.7	&	0.38	\\
La$_2$CuO$_4$(38)		&	9.61	&	1.37		&	3.7	&	6.1	&	4.6	&	6.2	&	0.79	&	140 (59)	&	1.2	&	3.9	&	0.36	\\
Hg1201(38)		                &	8.84	&	1.26		&	2.42	&	5.3	&	4.4	&	5.4	&	3.78	&	135 (76)	&	0.9	&	2.6	&	0.75	\\
Bi2212(78)				&	8.5	&	1.13		&	3.2	&	4.1	&	3.5	&	4.1	&	0.06	&	161 (76)	&	1.1	&	3.5	&	0.06	\\

\hline

Spectroscopy\\
\hline

CuO(54,65)	&	7.7	&	1.55		&	2.5	&	5	&	7.6	& 5.4	&	25.9	&	91 (79{$^\ast$})	&	0.8	&	2.6	&	1.72	\\
Sr$_2$CuO$_3$(62,63)		&	8.8	&	1.45		&	2.5	&	4.4	&	4.4	&	4.6	&	3.62	&	241 (71)	&	1.1	&	2.8	&	0.96	\\
 Sr$_2$CuO$_2$Cl$_2$(64)		&	8.8	&	1.5		&	3.5	&	4.4	&	6.0	&	4.6	&	6.45	&	130(60,61)	&	1.1	&	3.7	&	1.04	\\
 La$_2$CuO$_4$(65,66)		&	7.0	&	1.5		&	3.5	&	6.0	&	6.0	&	6.2	&	6.49	&	140 (59)	&	1.13	&	3.7	&	1.0	\\ 
 Nd$_2$CuO$_4$(67)		&	8.0	&	1.1		&	3.0	&	4.1	&	3.6	&	4.1	&	0.41	&	133 (72)	&	1.0	&	3.2	&	0.21	\\
 Pr$_2$CuO$_4$(67)		&	8.0	&	1.1		&	3.0	&	4.1	&	3.8	&	4.2	&	0.62	&	121 (72)	&	1.0	&	3.2	&	0.26	\\
 YBCO(53,55)		&	7.0	&	1.2		&	1.5	&	5.0	&	4.4	&	5.2	&	8.2	&	125 (73)	&	0.7	&	1.7	&	0.99	\\
  Bi2212(70)		&	7.7	&	1.5		&	3.5	&	6.0	&	5.6	&	6.1	&	4.3	&	161 (76)	&	1.2	&	3.7	&	0.88	\\
 																							
\hline																							
RIXS\\																							
\hline
 Bi2201(77)		&	10.2		&	1.35	&	3.9	&	5.9	&	4.5	&	5.9	&	0.34	&	153(76,77)	&	1.3	&	4.1	&	0.22	\\

\hline

\end{tabular}
	\end{center}
($^\ast$For CuO, the dominant
exchange interaction $J$ which couples antiferromagnetically is the one along the [10$\bar{1}$] chain direction and considered here, while the nearest neighbor spins exhibit a weak ferromagnetic coupling  (ref. 79).)
		\end{table} 

\begin{table}[t!]
	\begin{center}
	\caption{Renormalized electronic parameters $\tilde{U}$ and $\tilde{t}$ for cuprates in the one-band Hubbard model along with the ground state singlet weights c$_{21}$ (between the two Cu sites) and c$_{13}$ (between the Cu and O site). Optimization-1 uses  (${U}_d$, $t_{pd}$, ${\Delta_1}$, ${U}_{p1}$) and Optimization-2 uses  (${U}_d$, $t_{pd2}$, ${\Delta_2}$, ${U}_p$) from Table I to obtain corresponding $\tilde{U}$ and $\tilde{t}$.}

\begin{tabular}{|c|ccc|c|ccc|cc|}
\hline

 &      &    Optimization-1	&	& 	&		 & Optimization-2	 &	& &\\
Compound&      $\tilde{U}$	&	$\tilde{t}$	&	$\tilde{U}/\tilde{t}$	& 	$J$	&	$\tilde{U}$	 &	$\tilde{t}$	&	$\tilde{U}/\tilde{t}$	&	c$_{21}$	& c$_{13}$ \\

  	&	eV & eV & & ~~meV~~& eV   & eV & & & \\
\hline
																
Theory\\	
\hline
La$_2$CuO$_4$(34)		&	4.38		&	0.39	&	11.18	&140	&	3.68	&	0.36	&	10.26	&0.65&0.2\\
La$_2$CuO$_4$(35)		&	4.03		&	0.37	&	10.73	&140	&	3.7	&	0.36	&	10.28	&0.65&0.2\\
La$_2$CuO$_4$(36)		&	4.05		&	0.38	&	10.76	&140	&	3.81	&	0.37	&	10.42	&0.64&0.2\\
La$_2$CuO$_4$(38)		&	4.27		&	0.41	&	10.43	&140	&	4.05	&	0.4	&	10.16	&0.64&0.2\\
Hg1201(38)		                 &	3.93		&	0.36	&	10.79	&135	&	3.3	&	0.33	&	9.89	        &0.63&0.21\\
Bi2212(78)		&	3.35		&	0.37	&	9.13	        &161	&	3.34	&	0.37	&	9.11		&0.64	&0.2\\

\hline		

Spectroscopy\\
\hline
CuO(54,65)	                         &	4.39		&	0.32	&	13.87	&91	    &	3.09	&	0.27	&	11.62&0.64&0.2\\
Sr$_2$CuO$_3$(62,63)		&	3.79		&	0.48	&	7.94	        &241	    &	3.17	&	0.44	&	7.25&0.62&0.23\\
 Sr$_2$CuO$_2$Cl$_2$(64)	&	4.28		&	0.37	&	11.47	&130	    &	3.53	&	0.34	&	10.42&0.65&0.19\\
 La$_2$CuO$_4$(65,66)		&	3.96		&	0.37	&	10.64        &140    &	3.42	&	0.34	&	9.89	&0.64&0.2\\
 Nd$_2$CuO$_4$(67)		&	3.33		&	0.33	&	10.01	&133	    &	3.17	&	0.32	&	9.75	&0.64&0.2\\
 Pr$_2$CuO$_4$(67)		&	3.38		&	0.32	&	10.58	&121	    &	3.16	&	0.31	&	10.23&0.64&0.2\\
 YBCO(53,55)		                 &	3.49		&	0.33	&	10.56	&125	    &	2.61	&	0.29	&	9.14	&0.62&0.23\\
 Bi2212(70)		        		&	4.07		&	0.4	&	10.05	&161    &	3.59	&	0.38	&	9.44	&0.64&0.2\\

\hline																
RIXS \\																
\hline																
 Bi2201(77)		&	4.31		&	0.41	&	10.61	&153	&	4.18	&	0.4	&	10.45	&0.65	&0.2\\
 \hline
 
 \end{tabular}
	\end{center}
		\end{table} 

\begin{table}[t!]
	\begin{center}
	\caption{Examples of one-band model electronic parameters estimated independently from magnon dispersion in
neutron scattering, band dispersion in ARPES, and from {\it ab initio} theory.}
															
 \begin{tabular}{|c|ccc|c|}			
\hline
&&&&\\
Compound					&      ~~~$\tilde{U}$~~~	&	~~~$\tilde{t}$~~~	&	~~~$\tilde{U}/\tilde{t}$~~~	& 	Method	\\			
(ref. no)		  				&	eV 		& 	eV 		& 					& 	  \\
\hline	
&&&&\\						
 La$_2$CuO$_4$ (57)		       &	2.9		&	0.33	&	8.8		& 		Neutron scattering\\
 Sr$_2$CuO$_2$Cl$_2$ (58,59)      &	3.5		&	0.35	&	10.0		&         	ARPES\\
 La$_2$CuO$_4$ (38)		       &	5.0		&	0.48	&	10.4		& 		{\it ab initio} theory\\
 Hg1201 ~~(38)				       &	4.4		&	0.46	&	9.5		&         	{\it ab initio} theory\\
 &&&&\\
 \hline
\end{tabular}
	\end{center}
		\end{table}

\newpage
\twocolumngrid\
Given the differences in electronic parameters between the theoretical one-band versus the three-band models, and the corresponding experimental high-energy spectroscopies versus the low-energy magnon and band dispersion measurements, we felt it important to address a possible connection between them. In this study, we have found an equivalence by using the nearest neighbor Heisenberg exchange interaction $J$ obtained from neutron scattering and RIXS experiments\cite{Sr2CuO3,Coldea, Nd2CuO4,YBCO,Dean, Bi2212,Wang,Peng, CuO} as a bridge to connect electronic parameters known from experiment (high-energy spectroscopy, RIXS, neutron scattering) and theoretical estimates obtained from the one-band and three-band Hubbard models. The results show that the three-band Hubbard model parameters can be used to describe $J$ in terms of the well-known one-band Hubbard model form of $J = 4\tilde{t}^2/\tilde{U}$ with renormalized parameters $\tilde{U}$ and $\tilde{t}$.

We now summarize our main results.  We calculate $J$, the strength of the Heisenberg coupling between Cu moments in a Cu$_2$O cluster, using the downfolding method discussed by Koch \cite{Koch}. For several compounds neutron scattering data for $J$ and spectroscopic data for $U_d, ~U_p, ~\Delta$ and $t_{pd}$ are available. Directly using the spectroscopic parameter values in the downfolding expression for $J$ leads to deviations from the experimental $J$ values. We therefore use two estimation procedures, referred to as Optimization-1 and Optimization-2 in the following, to modify a subset (different in the two procedures) of the spectroscopic data, so that there is a good agreement with experimental $J$ values using the procedure described in Sec. \ref{model}. We identify effective parameters $\tilde{t}$ and $\tilde{U}$ so that the downfolding expression for $J$ becomes equal to  $4\tilde{t}^2/\tilde{U}$. The estimated values of $\tilde{t}$ and $\tilde{U}$ are found to be consistently smaller than $t_{pd}$ and $U_d$ in all cases, but lead to a larger $\tilde{U}/\tilde{t}$ in the one-band case compared to the $U_d/t_{pd}$ of the three-band case, in good agreement with theoretical estimates reported in the literature. 
The results indicate that stronger effective correlations, arising from a combination of $U_d, ~U_p, ~\Delta$ and $t_{pd}$ are hidden in the effective one-band Hubbard model. The study provides a consistent method to connect electronic parameters obtained from spectroscopy and the three-band model with effective parameters obtained from neutron scattering, ARPES  measurements and the one-band Hubbard model.

\section{Calculations}
\label{model}

We consider a Cu$_2$O cluster with site labels $i=1,2$ (for Cu(1) and Cu(2) atoms) and $i=3$ (for the O atom). The cluster hamiltonian is
\begin{eqnarray}
\label{eq:ham}
\hat{H} &=& \frac{\Delta}{2} (n_3 - n_1 - n_2 )  - t_{pd} \sum_{i\sigma} (d_{i\sigma}^{\dagger} p_{\sigma} + h.c.)    \nonumber \\
&+& U_d ( n_{1\uparrow} n_{1\downarrow} + n_{2\uparrow} n_{2\downarrow} ) + U_p n_{3\uparrow} n_{3\downarrow}, 
\end{eqnarray}
where $n_{i\sigma}^d = d_{i\sigma}^{\dagger} d_{i\sigma} ~ (i=1,2), ~n_{3\sigma} = p_{\sigma}^{\dagger} p_{\sigma}$ and $n_i = n_{i\uparrow} + n_{i\downarrow}$. Here, $d_{i\sigma}^{\dagger}$ creates a hole with a $z$ component of spin $\sigma = \pm 1/2$ in the Cu $d$ orbital at site $i ~(=1,2)$, and $p_{\sigma}^{\dagger}$ creates a hole with a $z$ component of spin $\sigma = \pm 1/2$ in the O $p$ orbital at the site located in between the two Cu sites. $\Delta$ is the charge-transfer energy; the parameters $U_p$ and $U_d$ are on-site Coulomb energies at the O and Cu sites, respectively; and finally $t_{pd}$ is the strength of hopping between neighboring O and Cu sites. 

In this work, we consider a filling fraction corresponding to undoped cuprates, in which the outermost $p$ orbital on the O site is filled with two electrons (i.e. empty in the hole picture), and the outermost $d$ orbital on the Cu site has one electron (i.e. one hole), in the absence of hopping. For our cluster with three atoms, this corresponds to a total occupancy of four electrons, or two holes. The two holes can be selected in three ways: both with $\sigma = -1/2$ (the ferromagnetic down case), both with $\sigma = 1/2$ (the ferromagnetic up case), and finally, one hole with $\sigma = 1/2$ and another with $\sigma = -1/2$ (the antiferromagnetic case). 

We will consider only the antiferromagnetic case henceforth. In this case, there are 9 basis states $| i, j \rangle, ~ i,j = 1,2,3$. In state $| i, j \rangle$, $i$ and $j$ are the Cu ($i,j = 1,2$) or O ($i,j = 3$) sites with up and down holes, respectively. In the basis $\{ | 1, 2 \rangle, ~ | 2, 1 \rangle, ~ | 1, 3 \rangle, ~ | 3, 1 \rangle, ~ | 2, 3 \rangle, ~ | 3, 2 \rangle, ~ | 1, 1 \rangle, ~ | 2, 2 \rangle, \\~ | 3, 3 \rangle \}$, the hamiltonian (\ref{eq:ham}) becomes a $9 \times 9$ matrix
\begin{gather}
\label{eq:hmat}
H =
  \begin{bmatrix}
H_{00}  &  H_{01} &  H_{02}    \\
H_{10}  &  H_{11} &  H_{12}    \\
H_{20}  &  H_{21} &  H_{22} 
   \end{bmatrix},
\end{gather}
in which the blocks are
\begin{gather}
H_{00} = -\Delta
  \begin{bmatrix}
   1 & 0 \\
   0 & 1 
   \end{bmatrix},
~H_{01} =
  \begin{bmatrix}
   t_{pd} & 0 & 0 & -t_{pd} \\
   0 & t_{pd} & -t_{pd} & 0 
   \end{bmatrix},  \nonumber \\
~H_{02} = O_{2 \times 3}, 
~H_{11} = O_{4 \times 4}, 
~H_{12} = 
  \begin{bmatrix}
   t_{pd} & 0 & t_{pd} \\
   -t_{pd} & 0 & -t_{pd} \\
   0 & t_{pd} & t_{pd} \\
   0 & -t_{pd} & -t_{pd}
   \end{bmatrix},  \nonumber \\  ~\mathrm{and} ~
~H_{22} = 
  \begin{bmatrix}
   U_d - \Delta & 0 & 0 \\
   0 & U_d - \Delta & 0 \\
   0 & 0 & U_p + \Delta
   \end{bmatrix}.
\end{gather}
In the above, $O_{m \times n}$ denotes an $m \times n$ matrix of zeros. We calculate the Heisenberg antiferromagnetic coupling $J$ between the Cu(1) and Cu(2) spins based on the downfolding fourth-order perturbation method described in Koch\cite{Koch} and Zurek\cite{zurek}. Accordingly, the effective hamiltonian is
\begin{eqnarray}
\label{eq:dfold}
\tilde{H} &=& H_{00} + H_{01} (\epsilon - (H_{11} + H_{12} (\epsilon - H_{22})^{-1} H_{21} )^{-1}))^{-1}. \nonumber \\
& \approx & H_{00} + H_{01} (\epsilon - H_{11})^{-1} H_{10} +  \nonumber \\
&& H_{01} (\epsilon - H_{11})^{-1} H_{12} (\epsilon - H_{22})^{-1} H_{21} (\epsilon - H_{11})^{-1} H_{10} \nonumber \\
\end{eqnarray}
where we have used the approximation $(A+B)^{-1} \approx A^{-1} (1 - B A^{-1})$. We now take $\epsilon = -\Delta$ and perform the matrix products above. The result is
\begin{gather}
\tilde{H} \approx  -\frac{2t_{pd}^2}{\Delta} 
\begin{bmatrix}
   1 & 0 \\
   0 & 1 
   \end{bmatrix}
-\frac{J}{2}
\begin{bmatrix}
   1 & -1 \\
   -1 & 1 
   \end{bmatrix}
\end{gather}
where the Heisenberg coupling is
\begin{eqnarray}
\label{eq:kochj}
J &=& 4 \frac{t_{pd}^4}{\Delta^2} \left[ \frac{1}{U_d} + \frac{1}{\Delta + U_p/2} \right].
\end{eqnarray}
The result we have obtained for $J$ using the downfolding approximation is the same as that reported in earlier studies \cite{Khomskii, zs1987, jbs2020} using a fourth-order perturbation theory. If we now define $\tilde{t}$ and $\tilde{U}$ using
\begin{equation}
\label{eq:ueff}
\tilde{t} = \frac{t_{pd}^2}{\Delta}, ~~~~\frac{1}{\tilde{U}} = \frac{1}{U_d} + \frac{1}{\Delta + U_p/2} 
\end{equation}
then $J=4\tilde{t}^2/\tilde{U}$. This is the expression for $J$ we would get if we used a one-band Hubbard model with a hopping strength $\tilde{t}$ and an on-site repulsion $\tilde{U} \gg \tilde{t}$, using second-order perturbation theory. In this sense, we consider $\tilde{U}$ and $\tilde{t}$ as the parameters of an effective one-band Hodel model corresponding to the model in (\ref{eq:ham}).

We find that using the spectroscopic values on the right-hand side and the experimental $J$ values on the left-hand side of equation (\ref{eq:kochj}) does not satisfy the equation with sufficient accuracy. We therefore modify, in an optimal manner described below, a subset of the spectroscopic parameter values so that the agreement is good.

We now describe two such simple optimization procedures. We define the parameter $R=\tilde{U}/\tilde{t}$ to obtain the parametric forms $\tilde{t} = RJ/4, ~\tilde{U} = R^2J/4$ for the effective parameters. These are the expressions that we use below for $\tilde{t}, ~ \tilde{U}$.

In the Optimization-1 procedure, we modify $(\Delta, ~ U_p)$. We solve the relations (\ref{eq:ueff}) to obtain $\Delta_1 (R) = t_{pd}^2/\tilde{t}$ and $U_{p1} (R)/2 = U_d \tilde{U}/(U_d - \tilde{U})-\Delta_1 (R)$. We then minimize a cost function $f_1 (R) = (\Delta_1 (R)-\Delta)^2 + (U_{p1} (R)-{U_p})^2$ with respect to $R$ to obtain the minimum $R^{\star}$, using the spectroscopic values for $U_d, ~U_p, ~\Delta$, $t_{pd}$ and neutron scattering values for $J$. This estimates $\Delta_1(R^{\star}), ~U_{p1}(R^{\star}), ~\tilde{U}(R^{\star}), ~\mathrm{and} ~\tilde{t}(R^{\star})$. 

In the Optimization-2 procedure, we modify $(\Delta, ~ t_{pd})$. We solve the relations (\ref{eq:ueff}) to obtain $t_{pd2}(R)^2 = \Delta_2 (R) \tilde{t}$ and $\Delta_2 (R) = U_d \tilde{U}/(U_d - \tilde{U})-U_p/2$. We then minimize a cost function
 $f_2 (R) = (\Delta_2 (R)-\Delta)^2 + |(t_{pd2}(R)^2-t_{pd}^2)|$ with respect to $R$ to obtain the minimum $R^{\star}$, using the spectroscopic values for $U_d, ~U_p, ~\Delta$, $t_{pd}$ and neutron scattering values for $J$. This estimates $\Delta_2(R^{\star}), ~t_{pd2}(R^{\star}), ~\tilde{U}(R^{\star}), ~\mathrm{and} ~\tilde{t}(R^{\star})$.

\section{Results and Discussion}
\label{results}

The results are summarized in Table I and Table II for a variety of CuO-based materials. Table I presents our estimates of the theoretical and spectroscopic three-band model parameters, while Table II presents our estimates of effective one-band parameters. Both tables contain results of the two optimization procedures that we discussed above.

In Table I, the columns 5-7 present results of Optimization-1 procedure: these are values of  $\Delta_1, U_{p1}$ and $f_1$; the columns 9-11 present results of Optimization-2 procedure: these are values of  $t_{pd2}, \Delta_2$ and $f_2$.  We can see from the cost function values that Optimization-2 is better than Optimization-1; this is also reflected in the greater closeness of $(t_{pd2}, \Delta_2)$ estimates to measured values than that of $(\Delta_1, U_{p1})$ estimates to measured values. Considering that $J$ depends on $t_{pd}^4$ in equation (\ref{eq:kochj}), it can be seen that a smaller spread in $t_{pd}$ across compounds provides a better description of parameter values; further since $\Delta$ and $t_{pd}$ are intimately related through the first relation in equation (\ref{eq:ueff}), it makes sense to optimize with respect to these two parameters as is done in Optimization-2. This has the result of reducing the spread in estimated $t_{pd2}$ values compared with reported spectroscopic $t_{pd}$ values. This also improves the estimates of $\Delta$ compared to $\Delta_1$ in Optimization-1. For these reasons, we can understand that Optimization-2 is better than not only Optimization-1, but also other possible optimization choices, namely $(\Delta, U_d)$, $(t_{pd}, U_p)$ and $(t_{pd}, U_d)$. We therefore do not present the results of these latter procedures.

Since $R=\tilde{U}/\tilde{t} \sim 10$ in most cases (see columns 4 and 8 in Table II), we can see that it makes sense to treat $\tilde{U}, ~\tilde{t}$ as effective one-band parameters, as is known from earlier work (Table III). We observe that $R \sim 10$ not only in cases where spectroscopic three-band parameter values are reported, but also for theoretical as well as RIXS three-band parameter values (see Table II).

Secondly, $\tilde{U}$ is roughly half of $U_d$ in almost all cases. This shows that the effective model is not a simple Cu $d$-band model, but possibly a more hybrid one involving Cu-$d$ and O-$(p_x, p_y)$ orbitals. To understand this better, we have looked at the nature of the ground state obtained by exactly diagonalizing the cluster hamiltonian (\ref{eq:ham}) for each compound in Table I using our Optimization-2 estimates of the three-band parameters. The ground state we obtain, $|G \rangle = \sum_{ij} c_{ij} |i,j \rangle$, always has the property $c_{12}=-c_{21}, c_{13}=-c_{31}=c_{23}=-c_{32}, c_{11}=c_{22}$ by symmetry, which shows that it is a singlet of Cu and O orbitals. We can thus completely characterize the ground state with the two distinct singlet weights $c_{21}$ and $c_{13}$ and the two distinct  hole double-occupancy weights $c_{11}$ and $c_{33}$. Since our focus is primarily on the singlet nature of $|G \rangle $, we present $c_{21}$ and $c_{13}$ in Table II (see columns 9 and 10). The singlet weights of the Cu-Cu and Cu-O sectors confirm that the ground state is a ZRS. The effective interaction $\tilde{U}$ is thus not between purely Cu-$d$ holes, but represents the hybrid ZR singlets, and is therefore significantly smaller than $U_d$. However, it must also be noted from Table I and Table II that $\tilde{U}/\tilde{t} \simeq 10$, satisfying the strong correlation condition in the effective one-band model.
In Table III, we list a few examples of one-band model electronic parameters estimated independently, from magnon dispersion in
neutron scattering, band dispersion in ARPES and from {\it ab initio} theory. It is clear that 
the values of $\tilde{U}/\tilde{t}$ in all the cases are close to the values in Table II and validates our analysis.

\begin{figure} 
\includegraphics[width=0.5\textwidth]{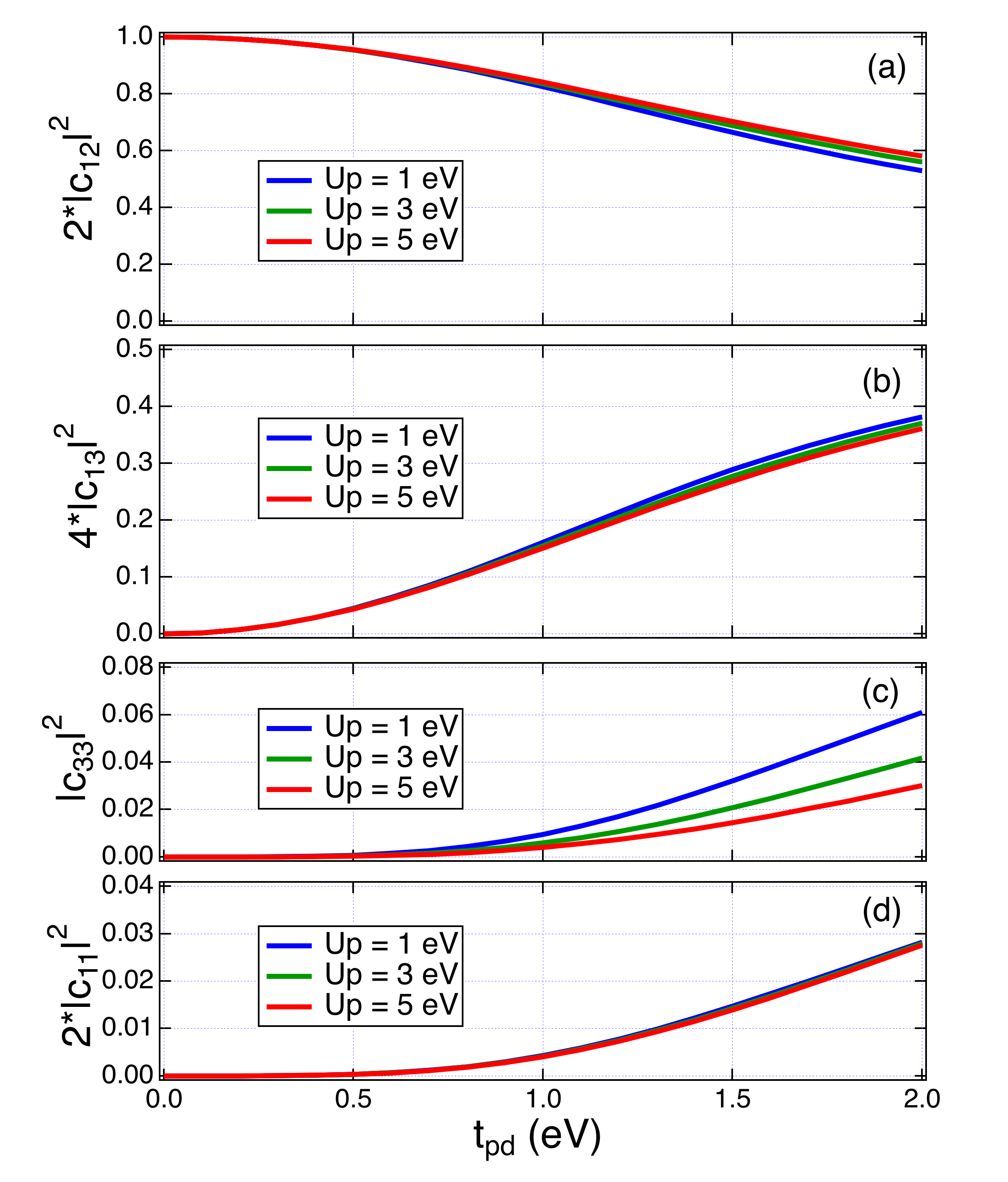}
\caption{(a-d) Plots of various calculated total weights $|c_{ij}|^2$ as a function of $t_{pd}$, for $U_d$=8 eV, $\Delta=3.3$ eV and $U_p$= 1, 3 and 5 eV. (a) shows the total Cu(1)-Cu(2) singlet weight 2$|c_{12}|^2$. (b) shows the total Cu(1)-O(3) singlet weight 4$|c_{13}|^2$. (c) and (d) show the on-site double occupancy weights 
$|2c_{11}|^2$ (total for the Cu(1) and Cu(2) sites) and $|c_{33}|^2$ for the O(3) site.}
\end{figure}

Finally, in Fig. 1(a-d) we present plots of various $|c_{ij}|^2$ as a function of $t_{pd}$, for 
$U_p = 1, 3$ and $5$ eV; we have fixed the values $U_d$=8 eV and $\Delta=3.3$ eV (=average $\Delta$ of values obtained by optimization 2 shown in Table I) in these plots. 
Fig. 1(a) shows that the pure Cu(1)-Cu(2) singlet weight 2$|c_{12}|^2$ $\sim$1 obtained for $t_{pd}$ = 0 systematically reduces in weight on increasing $t$. Simultaneously, the total Cu(1)-O(3) singlet weight 4$|c_{13}|^2$ increases systematically on increasing $t_{pd}$, indicating the role of Cu-O hybridization in forming the ZRS state for 
the cuprates. Thus, the O-$p$ orbital plays an increasingly important role on increasing $t_{pd}$ to $\sim1$ eV, typical  of the cuprates. Further, the on-site double occupancy weights $|c_{ii}|^2$, i = 1-3, are quite small (Fig. 1(c,d)). However, on increasing $U_p$ from 1  to 5 eV,  while there is hardly any change in the total double occupancy weight 2$|c_{11}|^2$ on the Cu sites, the double occupancy weight $|c_{33}|^2$ on the O site gets suppressed to nearly half its value for for $t_{pd} >1$ eV. This behavior of the O site double occupancy is closely related to the reduction of $J$ by $U_p$ according to Eq. (\ref{eq:kochj}). Thus, $U_p$ plays an important role in tuning the value of $J$, which is considered to be one of the most important parameters to achieve high-temperature superconductivity exhibited by the family of cuprates \cite{Keimer,Sr2CuO3,Coldea, Nd2CuO4,YBCO,Dean, Bi2212,Wang,Peng,Lipscombe,Braicovich,Dean2,Levy}.

\section{Concluding Remarks}
\label{conclude}

In this work, we have presented a data analysis of spectroscopic parameters and neutron-scattering parameters for a variety of cuprates based on a theoretical relationship between the parameters of a three-band model and an effective one-band Heisenberg antiferromagnetic coupling, using a cluster-model calculation. We have also performed an exact diagonalization of the cluster hamiltonian to understand the nature of the ground state.

Our analysis shows $\tilde{U} < U_d$ always. In addition to agreeing with estimates of $\tilde{U}$ from the one-band model applied to neutron scattering or ARPES experiments, this inequality is a direct consequence of equation  (\ref{eq:ueff}). 

$U_p$ is significant in magnitude, both in measurements and in our estimates, and is not small compared to $U_d$.
 While $U_p$ has been neglected in some studies on cuprates, we believe it is as important as $U_d$. Further, the second relation in equation (\ref{eq:ueff}) shows that the effective interaction $\tilde{U}$ is enhanced by $U_p$ and $\Delta$.

The ground-state singlet weights from our exact diagonalization show the importance of O moments and ZRS in the effective description. We also observe that the singlet weights change very little across the family of compounds, in spite of a variation in the three-band spectroscopic parameters that are used to calculate them. This holds for the ratio $\tilde{U}/\tilde{t}$ as well.

$\tilde{U}/\tilde{t} \sim 10$ in all cases, pointing to the effective one-band model being strongly correlated.

As to the spectroscopic values of the three-band parameters, it is generally believed that $\Delta$ and $t_{pd}$ measurements are less reliable than those of $U_d$ and $U_p$. Our estimation procedure Optimization-2 attempts to offer a reasonable description of the spectroscopic and neutron-scattering data by reducing the spread in the values of $\Delta$ and  $t_{pd}$ across the family of cuprates.

In conclusion, we have performed a perturbative and exact diagonalization study of a model of a Cu$_2$O cluster that connects  electronic parameters obtained from spectroscopy and the three-band model with values of $J$ obtained from scattering, band dispersion measurements and the effective one-band Hubbard model.  

\section{Acknowledgments} AF thanks JSPS KAKENHI (Grant Numbers JP19K03741 and JP22K03535) and the ``Program for Promoting Researches on the Supercomputer Fugaku" (Basic Science for Emergence and Functionality in Quantum Matter, JPMXP1020200104) from MEXT. AC thanks the National Science and Technology Council (NSTC) of the Republic of China, Taiwan for financially supporting this research under Contract No. MOST 111-2112-M-213-031.

\end{document}